\documentstyle[12pt]{article}

\begin{document}
\begin{flushright} {\bf CU-TP-889}
\end{flushright}
\vskip 20pt
\begin{center}
\begin{title}
\title{\Large\bf Running Coupling Effects in BFKL Evolution\footnote{This work is
supported in part by the Department of Energy under GRANT DE-FG02-94ER40819.}}
\end{title}
\vskip 20pt
\begin{author}
\author{Yuri V. Kovchegov  and A.H. Mueller}\\
{Physics Department, Columbia University\\
New York, New York 10027}
\end{author}
\end{center}
\begin{abstract} 
 We resum the recently calculated second order kernel of the BFKL
equation.  That kernel can be viewed as the sum of a conformally
invariant part and a running coupling part.  The conformally invariant
part leads to a corrected BFKL intercept as found earlier.  The
running coupling part of the kernel leads to a non-Regge term in the
energy dependence of high energy hard scattering, as well as a
$Q^2-$dependent intercept.
\end{abstract}

\section{Introduction}  

Recently, the calculation of the second order kernel for BFKL\cite{EAK,Yay}
evolution has been completed in two (partially) independent
calculations\cite{VSF,Cia}.  The second order kernel strongly modifies the
leading order result for the BFKL intercept at all but the smallest QCD
couplings.  While this large intercept correction does not significantly alter
our picture of BFKL dynamics, it suggests that even higher order corrections are
important leaving us without a reliable theoretical calculation of the
intercept which can, however, in principle be determined phenomenologically.

Of course, the corrected BFKL intercept governs single-scale high energy hard
scattering only after a resummation has been done to bring that intercept into
an exponential form.  It is the purpose of the present paper to do a more
complete resummation of the complete second order kernel for use at high
energies.

In Sec.2, we develop a formalism which allows one to iterate the complete
second order kernel an arbitrary number of times.  It turns out to be
convenient to separate the second order kernel into a conformally invariant
part and a running coupling part.  The resummation of the conformally invariant
part leads to the result given in Refs.3 and 4.  In Sec.3, after some
calculation, we resum the running coupling part of the kernel, and it also
exponentiates. The total result is given in (29) and (30).  The running coupling
part of the kernel has two effects.  It determines the scale of the leading
order intercept to be ${4N_c\ell n 2\over \pi} \alpha_s(qq^\prime)$ where $q $
and $q^\prime$ are the virtualities of the gluons which hook into the (compact)
particles for which the scattering is being calculated, as given in (1).  The
running coupling part of the kernel also introduces a non-Regge behavior in the
scattering, the ${D\over 3}[\alpha(\alpha_P-1)b]^2Y^3$ term in (30).  Indeed,
this non-Regge term is necessary in order to be able to view evolution in a
rapidity interval  $Y$  as built out of separate evolutions over regions $y$\ and
$Y-y$\ as indicated in (32).  Although this non-Regge term formally looks like a
next-to-next-to-leading term, in the exponent, it does not appear possible to
generate such a term from a third order kernel and so we believe the result
given in (30) is solid.  If one writes this term as $c \alpha^5 Y^3, c \approx
5$ for 3 flavors so that this term is reasonably small so long as
$\alpha$ is reasonably small and $Y$ is not inordinately large. 
Parametrically,  the Regge-type behavior in BFKL evolution begins to break
down when $Y \sim \alpha^{-5/3}$ a value where perturbation theory is still
valid, where the next-to-leading conformal corrections to the intercept are
(parametrically) small, but where unitarity corrections are expected to already
be large.  We note that E. Levin has previously arrived at $Y \leq
\alpha^{-5/3}$ as a criterion for the validity of the usual Regge-BFKL picture
of high energy scattering. (See Eq.82 of Ref.5.)

In Sec.4, we calculate the next-to-leading corrections to the anomalous
dimension and coefficient functions at large orders of perturbation theory.
The largest corrections to the coefficient function are determined completely
by the shift of the branch point in the angular momentum plane from $\omega =
\alpha_P-1$ to the position of the corrected BFKL intercept, while the largest
corrections to the anomalous dimension function are determined by the
appearance of a pole, $\gamma_{1\omega}={-\alpha(\alpha_P-1)b/4\over \omega -
(\alpha_P-1)}$, as indicated in (41).

Finally, in Sec.5, we investigate the limitations on using perturbation theory
to  calculate single-scale high energy hard scattering.  We do this by
examining the factorials which appear in perturbation theory and which determine
the region where the QCD asymptotic expansion can be used.  Eq.46 confirms the
result found in Ref.6 using a fixed coupling approach to the diffusion present
in BFKL evolution.

\section {Using the next-to-leading kernel to all orders}

The total cross section for the scattering of two compact colorless
particles (heavy onia) A and B can be written as

$$\sigma_{AB}(s) = \int {d^2q\over 2\pi q^2} \int {d^2q^\prime\over 2\pi
q^{\prime 2}} \Phi_A(\b{q})\Phi_B(\b{q}^\prime) \int_{a-i\infty}^{a+i\infty}
{d\omega\over 2\pi i}({s\over q q^\prime})^\omega G_\omega(\b{q},
\b{q}^\prime)\eqno(1)$$

\noindent where we use the notation of Ref.3.  $G_\omega$ obeys the equation

$$\omega G_\omega(\b{q}, \b{q}^\prime) = \delta(\b{q}-\b{q}^\prime) + \int
d^2\tilde{q}K(\b{q}, \tilde{\b q}) G_\omega(\tilde{\b q},
\b{q}^\prime)\eqno(2)$$

\noindent with

$$K(\b{q}, \b{q}^\prime) = K^{(1)}(\b{q}, \b{q}^\prime) + K^{(2)}(\b{q},
\b{q}^\prime)\eqno(3)$$

\noindent where the superscripts in (3) indicate the leading and 
next-to-leading BFKL kernels, respectively.  If $G_\omega^{(1)}$ represents
the leading BFKL result, the result of solving (2) taking only $K^{(1)}$ as
the kernel of that equation, then $G_\omega$ can be written as

$$G_\omega(\b{q}, \b{q}^\prime) = \sum_{N=0}^\infty G_{N\omega}(\b{q},
\b{q}^\prime)\eqno(4)$$

\noindent with

$$
G_{N\omega}=G_\omega^{(1)}[K^{(2)}G_\omega^{(1)}]^N\eqno(5)$$

\noindent where all of the products in (5) are understood to be convolutions as
indicated in (2).  It is convenient to write $G_\omega^{(1)}$ as

$$G_\omega^{(1)}(q,q^\prime) = {1\over \pi q q^\prime} \int {d\lambda\over 2\pi
i}({q\over q^\prime})^{2\lambda-1}{1\over \omega - {\alpha N_c\over
\pi}\chi(\lambda)}\eqno(6)$$

\noindent with $\chi(\lambda) = 2\psi(1) -
\psi(\lambda)-\psi(1-\lambda)$, and where we neglect the dependence on
$cos \phi = ({\b{q}\cdot \b{q}^\prime/{q}{q^\prime}})$ a dependence
which disappears when $\sigma(s)$ is evaluated at large $s.$ The
$\lambda$-integration in (6) runs parallel to the imaginary axis with
$Re\lambda = 1/2.$

It is useful to express the kernel $K^{(2)}$ as a conformally invariant part
and a ``running coupling'' part using the result\cite{VSF}

$$\int d^2q^\prime K^{(2)}(\b{q},\b{q}^\prime) ({q^\prime\over q})^{2\lambda-2}
= k^{(2)}(q)=k_{conf}(\lambda) + k_{rc}(\lambda,q)\eqno(7)$$

\noindent with

$$k_{conf}(\lambda)= - {1\over 4}({\alpha(\mu^2)N_c\over \pi})^2 c(\lambda)
\chi(\lambda)\eqno(8a)$$

\noindent and

$$k_{rc}(\lambda,q) = - {2\alpha^2(\mu^2)N_c\over \pi} b\chi(\lambda) \ell n\ 
q/\mu\eqno(8b)$$

\noindent where $c(\lambda)$ is defined in Ref.3, and where $b = {11 N_c-2N_f\over
12\pi}.$  Define

$$G_N(q,Y,q^\prime) = \int{d\omega\over 2\pi i} e^{\omega Y}
G_{N\omega}(q,q^\prime), \eqno(9)$$

\noindent along with similar definitions for $G$ and $G^{(1)}.$  Then,
substituting (6) in (5), using (9) as well as the identity

$$\int {d\omega\over 2\pi i}\  {e^{\omega Y}\over \omega - {\alpha N_c
\over \pi}\chi(\lambda)} \prod_{i=1}^N {1\over \omega - {\alpha N_c\over
\pi}\chi(\lambda_i)}$$
$$ = \int {d\omega\over 2\pi i} \prod_{i=1}^N {d\omega_i\over
2\pi i} dy_i{e^{\omega_i(y_i-y_{i-1})}\over \omega_i-{\alpha N_c\over
\pi}\chi(\lambda_i)}\ {e^{\omega(Y-y_N)}\over \omega - {\alpha N_c\over
\pi}\chi(\lambda)}, \eqno(10)$$

\noindent with $0 = y_0{\leq}y_1{\leq}y_2{\leq}\cdot \cdot
\cdot{\leq}y_N{\leq}Y,$ one finds

$$G_N(q,Y,q^\prime)=\int
G^{(1)}(q,Y-y_N,q_N)d^2q_N dy_Nk^{(2)}(q_N)$$
$$\prod_{i=1}^{N-1}
G^{(1)}(q_{i+1},y_{i+1}-y_i, q_i)d^2q_idy_ik^{(2)}(q_i)G^{(1)}
(q_1,y_1,q^\prime)\eqno(11)$$

\noindent where, in the saddle-point approximation,

$$G^{(1)}(q,y,q^\prime) = {e^{(\alpha_P-1)y}\over 2\pi qq^\prime{\sqrt{4\pi
Dy}}} exp[-{\ell n^2q/q^\prime\over 4Dy}]\eqno(12)$$

\noindent with $D={7\alpha(\mu^2)\over 2\pi} N_c\zeta(3)$ and $\alpha_P-1=
{4\alpha(\mu^2)N_c\over \pi} \ell n 2,$ where (12) is valid for large  $ {y}.$ 
Assuming $0 {<<}y_1{<<}y_2{<<}\cdot \cdot \cdot{<<}y_N{<<}Y$ in (11) one gets

$$G_N(q,Y,q^\prime) = {e^{(\alpha_P-1)Y}\over 2\pi
qq^\prime{\sqrt{4\pi DY}}}\ {1\over (4\pi D)^{N/2}}$$

$$\int{\prod_{i=1}^N}\ {dy_i du_i\over {\sqrt{y_i-y_{i-1}}}}
k^{(2)}(u_i){\sqrt{{Y\over Y-y_N}}}
exp[-{(u_i-u_{i-1})^2\over 4D(y_i-y_{i-1})}]
exp[-{(u-u_N)^2\over 4D(Y-y_N)}]\eqno(13)$$

\noindent where $u_i=\ell n\  q_i/\mu, u_0 = \ell n\  q^\prime/\mu$ and $u =
\ell n\  q/\mu.$ Now

$$k^{(2)}(u_i) = k_{conf} ({1\over 2}) + k_{rc}({1\over 2},u_i)\eqno(14)$$

\noindent with

$$k_{conf}({1\over 2}) = - {\alpha(\alpha_P-1)N_c\over 4\pi} c({1\over
2})\eqno(15a)$$

$$k_{rc}({1\over 2}, u_i) = - 2\alpha(\alpha_P-1) b\  u_i.\eqno(15b)$$

It is convenient to sum all orders of $k_{conf}$ for a given order of
$k_{rc}.$  Thus, one may write

$$G(q,Y,q^\prime) = {exp\{(\alpha_P-1)[1-{\alpha N_c\over 4\pi}c({1\over
2})]Y\}\over 2\pi qq^\prime{\sqrt{4\pi\ DY}}}\sum_{N=0}^\infty I_N\eqno(16)$$

\noindent with

$$I_N={1\over [4\pi D]^{N/2}} \int \prod_{i=1}^N{dy_i\over
{\sqrt{y_i-y_{i-1}}}} du_i k_{rc}(u_i){\sqrt{{Y\over
Y-y_N}}}exp[-{(u_i-u_{i-1})^2\over 4D(y_i-y_{i-1})}]$$
$$\cdot  exp[-{(u-u_N)^2\over
4D(Y-y_N)}].\eqno(17)$$

\noindent Choosing $Y = \ell n {s\over qq^\prime}$ and using (16) and (17) in
(1) gives the result of summing the leading and next-to-leading kernels to all
orders for the cross section $\sigma_{AB}.$  The first factor on the right-hand
side of (16) is the answer given in Refs.3 and 4.

\section{Evaluating the running coupling contribution}

In order to evaluate $I_N$ one can introduce a factor $exp\{\sum_{i=1}^N J_i
u_i\}$ in the integrand of (17), expressing each $u_i$ coming from
$k_{rc}(u_i)$ as ${\partial\over \partial J_i}$ acting on the integral, and then
setting the $J_i=0$ at the end of the calculation

$$I_N=[-2\alpha(\alpha_P-1)b]^N {1\over (4\pi D)^{N/2}} \prod_{i=1}^N
{\partial\over \partial J_i}\int {dy_i\over
{\sqrt{y_i-y_{i-1}}}}du_i{\sqrt{{Y\over Y-y_N}}}$$

$$\cdot exp\{-{(u_i-u_{i-1})^2\over 4D(y_i-y_{i-1})} + J_iu_i\}
exp\{-{(u-u_N)\over 4D(Y-y_N)}\}\vert_{_{J=0}}.\eqno(18)$$

\noindent It is useful to make the change of variables

$$z_0 = {u_0\over {\sqrt{Y}}} {1\over {\sqrt{4D}}}$$
$$z_i = {y_{i+1}(u_i-u_0)-y_i(u_{i+1}-u_0)\over
{\sqrt{y_iy_{i+1}(y_{i+1}-y_i)}}}\  {1\over {\sqrt{4D}}}, i = 1,2,\cdot \cdot
\cdot, N\eqno(19)$$

$$z = {u\over {\sqrt{Y}}}\ {1\over {\sqrt{4D}}}$$

\noindent where $y_{N+1}=Y, u_{N+1}=u.$  The $u_i$ are expressed in terms of
the $z_i's$ by

$$u_i-u_0 = y_i\left[\sum_{j=i}^N z_j{\sqrt{{y_{j+1}-y_j\over y_jy_{j+1}}}}
+{z-z_0\over {\sqrt{Y}}}\right]{\sqrt{4D}}\eqno(20a)$$

$$u-u_0 = Y{z-z_0\over {\sqrt{Y}}}{\sqrt{4D}}.\eqno(20b)$$

\noindent After rescaling $J_i\to {\sqrt{4D}}\  J_i$ we have

$$I_N=[-2\alpha(\alpha_P-1)b]^N exp\{-{1\over 4DY} \ell n^2{q\over
q^\prime}\}({4D\over \pi})^{N/2}$$
$$\left[\prod_{i=1}^N
{\partial\over \partial J_i} \int dy_i dz_i exp\{-z_i^2+
z_i{\sqrt{{y_{i+1}-y_i\over y_{i+1}y_i}}} \sum_{j=1}^i J_jy_j\}\right]$$
$$\cdot
exp\{{1\over {\sqrt{4D}}}\ell n {q^\prime\over \mu} (J_1+\cdot \cdot \cdot +
J_N) + {1\over {\sqrt{4D}}Y}\ell n {q\over q^\prime} (J_1y_1+ \cdot \cdot \cdot +
J_Ny_N)\}\vert_{_{J=0}}\eqno(21)$$

\noindent where, as always, the $y$-integration is ordered
$0{\leq}y_1{\leq}y_2{\leq}\cdot\cdot\cdot{\leq}y_N{\leq}Y$ and where it is
understood that all the ${\partial\over \partial J_i}$ factors are to be put to
the left of all the $J_i$ terms in the various exponents in (21).  After
performing the integration over $dz_i$ and after some simple algebra formula
(21) becomes

$$I_N=[-2\alpha(\alpha_P-1)b]^N exp\{- {1\over 4DY} \ell n^2 {q\over
q^\prime}\} (4D)^{N/2}$$
$$\left[\prod_{i=1}^N{\partial\over \partial J_i} \int dy_i
exp\{{1\over 2} \sum_{j=i+1}^N J_iJ_j{y_i(Y-y_j)\over Y}\}\right]$$

$$\cdot exp\{{1\over {\sqrt{4D}}}\ell n {q^\prime\over \mu}(J_1+\cdot \cdot
\cdot + J_N) + {1\over {\sqrt{4D}}Y} \ell n {q\over q^\prime} (J_1y_1+ \cdot
\cdot \cdot + J_Ny_N)\}\vert_{_{J=0}}.\eqno(22)$$

\noindent Moving the terms linear in  J  in the exponent to the left of the
${\partial\over \partial J_i}$ factors gives

$$I_N=[-2\alpha(\alpha_P-1)b]^N exp\{-{1\over4DY} \ell n^2{q\over q^\prime}\}
(4D)^{N/2}\prod_{i=1}^N$$
$$\int dy_i\left[{1\over {\sqrt{4D}}}\ell n {q^\prime\over
\mu} + {1\over {\sqrt{4D}}}{y_i\over Y} \ell n {q\over q^\prime} +
{\partial\over \partial J_i}\right]exp\{{1\over 2} \sum_{j=i+1}^N J_iJ_j
{y_i(Y-y_j)\over Y}\}\vert_{_{J=0}}.\eqno(23)$$

Now consider the expression

$$P_k=\prod_{i=1}^k \int dy_i {\partial\over \partial J_i} exp\{{1\over 2}
\sum_{j=i+1}^k J_iJ_j{y_i(Y-y_j)\over Y}\}\vert_{_{J=0}}.\eqno(24)$$

\noindent $P_k$ is non-zero only for even k  in which case

$$P_k = {1\over 2^{k\over 2}}\ {1\over ({k\over 2})!}\  \int dy_1\cdot \cdot \cdot
dy_k {\sum\atop Perm}\ {\prod\atop {i,j\atop i<j}}\  {y_i(Y-y_j)\over
Y}                 
\eqno(25)$$

\noindent where the sum goes over all permutations of the pairs $i$  and  $j.$ 
Eq.25 is readily evaluated as

$$P_k = {1\over 2^{k/2}}\  {1\over ({k\over 2})!}\  \left[\int_0^Y dy_2
\int_0^{y_2}dy_1{y_1(Y-y_2)\over Y}\right]^{k/2} = {1\over ({k\over
2})!}\left[{Y^3\over 2\cdot 4!}\right]^{k/2}.\eqno(26)$$

\noindent Rewriting (23) as

$$I_N=[-2\alpha(\alpha_P-1)b]^N exp\{-{1\over 4DY} \ell n^2{q\over
q^\prime}\}(4D)^{k/2}\sum_{k=0}^N P_k{1\over (N-k)!}
[{Y\over 2}\ell n{qq^\prime\over \mu^2}]^{N-k}\eqno(27)$$

\noindent we obtain

$$I_N=exp\{-{1\over 4DY} \ell n^2 {q\over q^\prime}\} \sum_{k=0}^N {1\over
(N-k)!}[-\alpha(\alpha_P-1)b\  Y \ell n {qq^\prime\over \mu^2}]^{N-k}\cdot
{1\over ({k\over 2})!}$$
$$\cdot[{1\over 3} (\alpha(\alpha_P-1)b)^2 D Y^3]^{k/2}\eqno(28)$$

\noindent where the sum goes over even values of $k.$  Using (28) in (16) gives

$$G(q,Y,q^\prime) = {e^{E(q,Y,q^\prime)}\over 2\pi\ qq^\prime{\sqrt{4\pi
DY}}}\eqno(29)$$

\noindent with

$$E(q,Y,q^\prime) = (\alpha_P(qq^\prime)-1)(1-{\alpha N_c\over 4\pi}c({1\over
2}))Y + {D\over 3} (\alpha(\alpha_P-1)b)^2Y^3-{\ell n^2q/q^\prime\over
4DY}\eqno(30)$$

\noindent where we have used

$$(\alpha_P(\mu^2)-1)(1-\alpha(\mu^2)b(u_0+ u)) =
\alpha_P(qq^\prime)-1.\eqno(31)$$
\clearpage

The $-(\alpha_P-1){\alpha N_c\over 4\pi}c({1\over 2})$ term in (31) gives the
next-to-leading correction to the BFKL intercept\cite{VSF,Cia}.  This term
comes from the conformally invariant part of the next-to-leading kernel, as
given by (8a) and (15a).  The running coupling part of the next-to-leading
kernel gives the scale of the leading BFKL intercept, the
$\alpha_P(qq^\prime)-1={4N_c\over \pi}\ell n 2 \alpha(qq^\prime)$ term in (30),
and also the non-Regge term, the $\alpha^5Y^3$ term in (30).  The appearance
of a $Y^3$ term in the exponent signals a breakdown\cite{Lev} of the usual
picture of high energy scattering where powers of $Y$ and the factorial
denominators associated with them come only from a strongly ordered region of
longitudinal phase space.  The $\alpha^5Y^3$ term in (30) is, however, purely
perturbative and the overall coefficient is reasonably small, ${D\over
3}[\alpha(\alpha_P-1)b]^2Y^3 \approx 5\alpha^5Y^3,$ so that this term is likely
not too important for present phenomenology.  We note that the expression (29),
with 
$E$  given in (30), obeys the consistency condition

$$G(q,Y,q^\prime) = \int d^{2}kG(q,y,k) G(k,Y-y,q^\prime),\eqno(32)$$

\noindent and that the $\alpha^5Y^3$ term in $E$ is crucial for (32) to hold. 
Indeed, one can view (32) as requiring the $\alpha^5Y^3$ term, exactly as given in
(30), once one has arrived at the scale $qq^\prime$ as the appropriate scale for
the leading BFKL intercept.

\section{The anomalous dimension and coefficient functions
at  large orders[7-10]}

Certain higher order corrections to the gluon anomalous dimension have already
been calculated\cite{VSF}.  Here we do a much simpler calculation.  We
shall  calculate the dominant large order terms at next-to-leading level for both
the coefficient function and the anomalous dimension function.  These large order
terms are insensitive to the scheme in which the energy scale is introduced.

To determine the coefficient function $C_\omega(\alpha)$ we take $q=q^\prime =
\mu$ in (29) and write

$$\pi qq^\prime G(q,Y,q^\prime)\vert_{_{q=q^\prime=\mu}}=\int {d\omega\over 2\pi
i} C_\omega e^{\omega Y}.\eqno(33)$$

\noindent Keeping only the first next-to-leading order term one gets

$${e^{(\alpha_P-1)Y}\over 2{\sqrt{4\pi DY}}}(1-{\alpha(\alpha_P-1)N_c\over 4\pi}
c({1\over 2})Y) = \int {d\omega\over 2\pi i} (C_{0,\omega}+ C_{1,\omega})
e^{\omega Y}\eqno(34)$$

\noindent where

$$C_{0,\omega} = \sum_{N=0}^\infty ({\alpha N_c\over \pi \omega})^N {1\over
\omega} C_0^{(N)},\ C_{1,\omega} = \sum_{N=0}^\infty ({\alpha N_c\over
\pi\omega})^NC_1^{(N)}.\eqno(35)$$

\noindent The $C_0^{(N)}$ are given by \cite{AHM}

$$C_0^{(N)}\widetilde{{_{_N\rightarrow\infty}}} {1\over
{\sqrt{N}}}({\alpha_P-1\over
\alpha N_c/\pi})^{N+{1\over 2}}{1\over {\sqrt{56\pi \zeta(3)}}}\eqno(36a)$$

\noindent while a simple calculation, using (34), leads to 

$$C_1^{(N)} \widetilde{{_{N\rightarrow \infty}}} - {Nc({1\over 2})\over 4}
C_0^{(N-1)}. \eqno(36b)$$

In order to determine the anomalous dimension function at
next-to-leading level one can take $q=\mu, q^\prime = Q$ and evaluate
(29) to first order in $\ell n\ Q^2/\mu^2.$ One finds

$${-e^{(\alpha_P-1)Y}\over 2{\sqrt{4\pi DY}}} \alpha(\alpha_P-1) bY\ell n Q/\mu
= \int {d\omega\over 2\pi i}[C_\omega(\gamma_\omega-{1\over 2})]_{_1} \ell n\ 
Q^2/\mu^2 e^{\omega Y}\eqno(37)$$

\noindent where $[C_\omega(\gamma_\omega - {1\over 2})]_{_1}$ indicates the
next-to-leading corrections to the product $C_\omega(\gamma_\omega-{1\over
2}).$  That is

$$[C_\omega(\gamma_\omega-{1\over 2})]_{_1}=\sum_{N=0}^\infty [C(\gamma - 
{1\over
2})]_{_1}^{(N)}({\alpha N_c\over \pi\omega})^N.\eqno(38)$$

\noindent Using (37) it is straightforward to determine

$$[C(\gamma - {1\over 2})]_{_1}^{(N)} = - ({\alpha_P-1\over \alpha
N_c/\pi})^{N-{1\over 2}}{b\pi\over 2 N_c} {\sqrt{N\over
56\zeta(3)\pi}}\eqno(39)$$

\noindent which, using (35) and

$$\gamma_{0,\omega} = \sum_{N=0}^\infty \gamma_0^{(N)}({\alpha N_c\over
\pi\omega})^N,\  \gamma_{1,\omega} = \sum_{N=1}^\infty \gamma_1^{(N)} {\alpha
N_c\over \pi} ({\alpha N_c\over \pi\omega})^{N-1}\eqno(40)$$

\noindent along with\cite{AHM} $\gamma_0^{(N)} = {1\over N} C_0^{(N)},$ leads to

$$\gamma_1^{(N)} \widetilde{{{_N\rightarrow \infty}}} - {\pi b\over 4N_c} ({\alpha
N_c/\pi\over \alpha_P-1})^{3/2} {\sqrt{56\pi \zeta(3)}} N^{3/2}\gamma_0^{(N)} =
- {b\pi\over 4N_c}({\alpha_P-1\over \alpha N_c/\pi})^{N-1}.\eqno(41)$$

The correction for $C_1^{(N)}$ given in (36b) simply corresponds to a shift in
the branch point in $C_\omega$ from  $C_\omega
=[16D(\omega-(\alpha_P-1))]^{-{1\over 2}}$

\noindent to  $C_\omega =\{16D[\omega-(\alpha_P-1)(1-{\alpha N_c\over 4\pi}
c({1\over 2}))]\}^{-{1\over 2}}.$   The $N^{3/2}$ factor in (41) indicates the
start of the appearance of a non-Regge term.  The fact that $C_1^{(N)}/C_0^{(N)}$
and $\gamma_1^{(N)}/\gamma_0^{(N)}$ are large\cite{Bal} at large $N$ does not in
itself signal a breakdown of perturbation theory since these terms are known
exactly and, at least in the case of $\gamma_1^{(N)},$ lead to corrections which
are not particularly large.

\flushleft\section{Limitations on the use of perturbation theory}

Because of the diffusion inherent in BFKL dynamics it is clear that the
perturbative approach  to high energy single-scale short distance behavior must
breakdown at sufficiently high energy.  Estimates of the energies at which that
breakdown occurs were given in Ref.6.  We are now in a position to see this
breakdown in a more detailed way than has been done previously. The idea is
simple. Perturbation theory should itself indicate when it is breaking down by
generating factorial terms which indicate where the asymptotic expansion, the
perturbative expansion, is reliable.

Suppose we take the $N=1$ term in (5), but now instead of taking the
running coupling correction to be given by (8b) we work to arbitrary
order in running coupling corrections

$$\tilde{k}_{rc}(\lambda,q) = {\alpha(\mu^2)N_c\over \pi} \chi(\lambda)
\sum_{n=1}^\infty(-2b\alpha(\mu^2)\ell n\  q/\mu)^n.\eqno(42)$$

\noindent For simplicity we take $q=q^\prime = \mu$ in evaluating G.  Thus, we
replace $I_1$ in (16), (18), and (21) with

$$\tilde{I}_1 = \sum_{n=1}^\infty I_{1,n}\eqno(43)$$

\noindent where

$$I_{1,n} = {[-2\alpha(\mu^2)b]^n(\alpha_P-1)\over {\sqrt{4\pi D}}}
\int_0^Y dy_1\int_{-\infty}^\infty dz_1 e^{-{z_1^2\over
4D}}\left[y_1z_1{\sqrt{{Y-y_1}\over y_1Y}}\, \right]^n.\eqno(44)$$

\noindent A simple calculation leads to

$$I_{1,n} = {(\alpha_P-1)Yn\over 2(n+1)} (4\alpha^2b^2DY)^{n/2}\Gamma({n\over
2})\eqno(45)$$

\noindent for $n$ even, and $I_{1,n}=0$ for $n$ odd.  Of course the
sum indicated in (43) does not converge, and this is not surprising.
So long as $4b^2\alpha^2DY {<<} 1, I_{1,n}$ is small for small values
of $n$ and the divergence of the sum in (43) is a standard renormalon
problem. One can keep those terms in $n$ so long as ${I_{1,n+1}\over
I_{1,n}} \leq 1,$ with higher orders in $n$ being discarded.  However,
when $4\alpha^2b^2DY \approx 1$ we are not allowed to safely use any
of the terms in the perturbation series given in (45), and thus the
whole perturbative approach breaks down.  The criterion for being able
to use perturbation theory without too much contamination from
infrared regions then is

$$4\alpha^2Db^2Y {\leq} 1\eqno(46a)$$

\noindent or

$$Y {\leq} {\pi\over 14N_c\zeta(3)b^2} {1\over \alpha^3(\mu^2)}\eqno(46b)$$

\noindent exactly as found in Ref.6.  $\mu$ characterizes the scale of the
single-scale hard process in question.
\clearpage
\noindent{\bf Acknowledgements}

One of the authors (AM) acknowledges very helpful discussions with J. Bartels,
M. Ciafaloni, V.Fadin and L. Lipatov concerning higher order corrections to BFKL
evolution.  Yu.K. wishes to thank E. Levin for a number of stimulating
and informative discussions about running coupling effects in BFKL evolution.

\end{document}